# Friedel oscillations and helium bubble ordering in molybdenum

W. T. Geng[*] and Q. Zhan

*School of Materials Science and Engineering, University of Science and Technology Beijing, Beijing 100083, China.*

Helium ions implanted into metals can evolve into ordered bubbles isomorphic to the host lattice. Long-range elastic interaction is generally believed to drive the formation of bubble superlattice, but little is known about the thermodynamics at the very initial stage. Our first-principles calculations demonstrate that in molybdenum, Friedel oscillations induced by individual helium generate both potential barriers and wells for helium clustering at short He-He distances. Such repulsion and attraction at high concentration provide a thermodynamic diving force to assist lining up randomly distributed He atoms into ordered bubbles. Friedel oscillations might have general impact on solute-solute interactions in alloys.

---

[*] geng@ustb.edu.cn

When a metal is irradiated with neutrons or heavier particles at elevated temperatures, abundant vacancies emerge and agglomerate into microcavities known as voids, or bubbles if they are filled with gas. It was discovered by Evans half century ago that such voids in Mo form a nearly perfect body-centered-cubic (bcc) lattice structure, isomorphic to the matrix crystal.[1] Soon after, bcc superlattice of He bubbles were also found in Mo,[2] and extensive investigations have revealed self-organization of isomorphic superlattice of voids and He bubbles under high-dose irradiation in many other metals, with bcc, face-centered-cubic (fcc), or close-packed-hexagonal (hcp) lattices.[3,4,5] Understanding the physical origin of this nanostructure self-organization is both important in fundamental condensed matter physics and crucial from technological perspectives because knowledge of the dependence of the microstructure on the irradiation and material conditions can be helpful in the design of radiation-resistant materials in nuclear reactors.

Various models have been proposed to explain the formation of void and bubble superlattices. The early models proposed by Tewary and Bullough,[6] Stoneham,[7] and Wills[8] considering only elastic interactions between voids, can reproduce the existence of an energy minimum at a particular void-void distance and give somewhat reasonable ratios between this distance and void radius in Mo, but leave the dynamic ordering process totally unknown. Other authors explored the concept of phase transition in nonequilibrium conditions,[9] the idea of anisotropic diffusion of self-interstitial-atoms,[10,4] or theory of non-linear dynamics for dissipative systems.[11] More recent models take into account both elasticity and dynamic diffusion process based on the phase-field method, and obtain better agreement with experimental observations on the growth and ordering of voids and bubbles in different metals. [12,13,14]

It has been well established that in the early stages of the superlattice formation, small voids or bubbles are distributed randomly in the host metal.[15,4,11,14] It appears that the initially disordered small voids move around in a Brownian motion until a small group of them happen to reach a correct configuration to nucleate a superlattice at an appropriate lattice parameter which minimize the free energy. However, it has to be stressed that fundamental information about the stability of individual solute He atoms, at the very initial stage of bubble formation, is still missing.

Due to the strong He-vacancy binding[16,17] and the abundancy of vacancies under irradiation conditions, the implanted He may immediately take the substitutional positions, presumably in a disordered manner. Since the 1*s* electrons in a He atom are deep-level and paired, He experiences Pauli repulsion in electron gas and cannot form chemical bond with any other elements.[18] Therefore, substitutional He atoms have strong tendency to merge into clusters and finally into bubbles to reduce the He-matrix interfacial energy. Then, an important question arises: Why individual solute He atoms have the tendency to stay away from each other before merging into clusters? Apparently, the underlying repulsive force play a crucial role in the formation of a dense and randomly positioned small bubbles before they merge and reassemble into a periodic array of larger bubbles.

Such a He-He repulsive force has a long-range nature since the two He atoms are separated by at least one host atom in the matrix lattice before forming a He-He pair. One might be tempted to assume that it must be originated from the elastic interaction between point-defects. However, it is known from first-principles calculations that substitutional He in W, which has a lattice constant very close to Mo, has an effective atomic size nearly the same as W atoms (difference in radius < 0.01Å).[19] This means elastic interaction should be very weak between individual He atoms, before they agglomerate into bubbles. In trying to tackle this puzzle, we recall two facts: (i), He as an impurity will produce Friedel oscillations[20] of the electron density in the host metal,[21] and (ii), He almost always prefers a low electron-density region in metals.[18] The two features inspire us to conjecture that it might be the effect of Friedel oscillations that give rise to the sizable long-range He-He interactions in metals.

To verify this hypothesis, we have employed first-principles density functional theory (DFT) calculations to evaluate the He-He interaction in Mo at different concentrations, in terms of solution energy variation. To keep computation affordable, the individual substitutional He atoms are put in symmetric lattice sites in Mo, forming a superlattice based the original lattice of Mo. By itself, the superlattice of implanted He atoms is simple cubic, bcc, or fcc. Note that a substitutional He can be viewed as a combination of He and vacancy, which is superabundant under irradiation. These models are illustrated in Fig. 1, taking (2×2×2) supercells as an example.

**Fig. 1. Computational model for ordered substitutional He in Mo, forming an fcc (a), bcc (b), or simple cubic (c) superlattice. The (2×2×2) host supercells are taken as an example.**

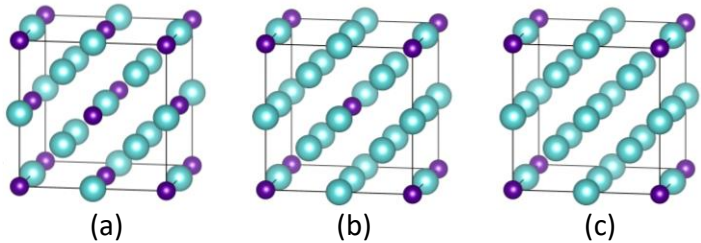

(a)　　　　　(b)　　　　　(c)

We have performed the DFT calculations using the Vienna Ab initio Simulation Package [22]. The electron-ion interaction was treated using the projector augmented wave (PAW) method [23]. The exchange correlation between electrons was described with generalized gradient approximation (GGA) in the Perdew-Burke-Ernzerhof (PBE) form [24]. We used an energy cutoff of 500 eV for the plane wave basis set for all systems to ensure equal footing. To calculate the solution energies of ordered He at different concentrations, we have employed supercells of bcc Mo from (2×2×2) to (7×7×7), modelling concentration from 25 at% to 0.6 at% for an array of He in fcc superlattice, from 12.5 at% to 0.3 at% for a bcc superlattice. For a simple cubic He superlattice, we included the (1×1×1) Mo cell so the He concentration varies from 50 at% to 0.15 at%. The Brillouin-zone integration was performed within the Monkhorst-Pack scheme using a dense $k$ mesh of (24×24×24), (12×12×12), (8×8×8), (6×6×6), (5×5×5), (4×4×4), (4×4×4) for Mo supercells, respectively. To investigate the short-range He-He interaction at low concentration, we put two He atoms in a (7×7×7) supercell along the <111> direction and calculate the energy dependence on the He-He distance. For all supercells, we started from the configurations in which He take substitutional positions and all the matrix atoms are in prefect bulk sites and optimize both internal coordinates and supercell dimensions. The energy relaxation for each supercell was continued until the total energies were converged to less than $1\times10^{-4}$ eV/cell. We note that the total energy of the initial configuration is the one before the He-He elastic interactions taking effect.

The solution energy of He in Mo, $\Delta H_s(\text{He})$, can be obtained via

$$\Delta H_s(\text{He}) = \frac{1}{n}[E(\text{Mo}_{2m^3-n}\text{He}_n) - (1 - \frac{n}{2m^3}) \times E(\text{Mo}_{2m^3}) - nE(\text{He})] \quad (1)$$

where $E(\text{Mo}_{2m^3-n}\text{He}_n)$ is the total energy of the $(m \times m \times m)$ supercell containing $2m^3-n$ Mo and $n$ He atoms, $E(\text{Mo}_{2m^3})$ is the total energy of the clean $(m \times m \times m)$, $E(\text{He})$ is the total energy of an isolated He atom, which is set to zero, and $n$=4, 2, or 1 for an fcc, bcc, or simple cubic superlattice of He in Mo. We display in Fig.2 the calculated formation energy of substitutional He atoms in Mo. For the case of isomorphic distribution, we also show the formation energy of He in *frozen* configurations (open symbols), in addition to that in fully relaxed configurations (solid symbols). By *frozen* we mean all atoms in a supercell are in the ideal lattice sites, and the contribution of He-He elastic interaction to the total energy is absent.

**Fig. 2. The formation energy of substitutional He in Mo supercells.**

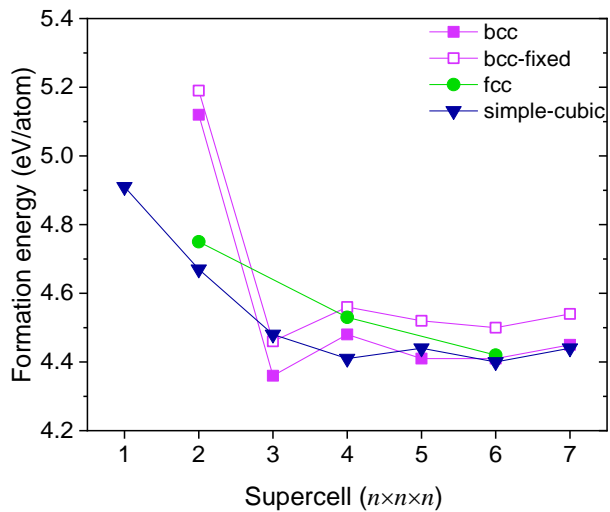

One striking feature shown in Fig. 2 is that the formation energy of He takes the highest value in the smallest supercell. This indicates that before collectively merging into clusters upon which there is significant energy release, He atoms experience a repulsion from each other at high concentration. Such a barrier changes very little from 0.76 to 0.73 eV upon geometry optimization when elastic interactions take effect. It can therefore be concluded there must be a driving force other than He-He elastic interactions responsible for the oscillation of the solution energy against concentration. Another interesting finding is that the solution energy of He reaches a local minimum before getting to the barrier, in the isomorphic bcc structure. By comparison, the solution energy of non-isomorphic array of He generally increases from low to high. The nearest-neighbor He-Mo distance in the configuration of one He in (7×7×7) Mo supercell is 2.728 Å, only 0.005Å larger than the Mo-Mo bond length in bulk, and hence a marginal size effect. It is for this reason that the difference in the formation energy of He in the *frozen* supercells and the fully optimized ones does not vary remarkably with concentration, indicating that the He-He elastic interactions are very weak. It has to be stressed that in the absence of elastic interactions, Friedel oscillations alone can also give rise to rearrangement of atomic positions, as is evidenced in relaxation of free surfaces.[25]

**Fig. 3. The Friedel oscillations of charge density along the [100], [110], and [111] directions in Mo upon replacing one Mo with a helium in a (7×7×7) supercell. Distance from the He atom is in scale of the nearest-neighbor Mo-Mo distance $d_{nn}$=2.723 Å. (b) Fitting of oscillations in [111] direction. The insets are closeups for large *r*.**

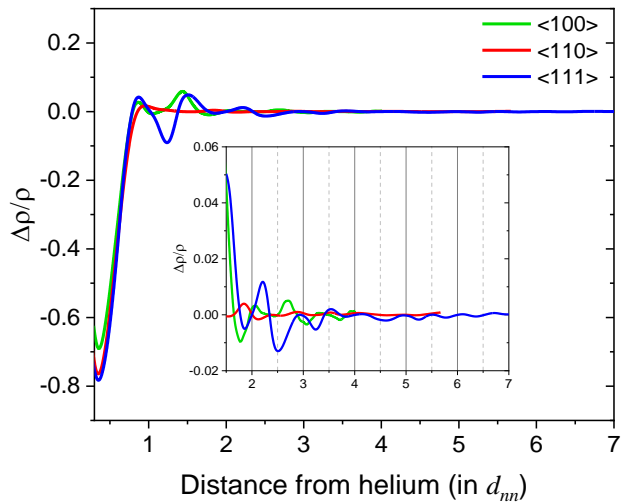

Friedel oscillations are spatial modulations of the electron density in a metal whenever a local disturbance occur.[20] Static Friedel oscillations have the form of a standing-wave perturbation and fall off with distance from the disturbance. Near a point impurity, the relative change of electron density, $\frac{\Delta\rho}{\rho}$, decays with distance, $r$, in a jellium model as[21]

$$\frac{\Delta\rho}{\rho} \sim \cos(2k_f r - \varphi) \times \frac{1}{r^3}, \tag{2}$$

where $k_f$ is the Fermi wavevector of the electron gas in the host metal and $\varphi$ is the phase shift accounting for the electron transfer between the impurity and the host. The discrete nature of the crystal lattice means Friedel oscillations are orientation-dependent in real metals, following the Fermi wavevector variation.[26] Keeping in mind that He prefers a low electron density environment when substituting for a matrix atom, we are more interested in the electron density variation along atomic chains containing the He atom. We choose three low-index directions, <100>, <110>, and <111>, and calculated the relative change of charge density along these three directions upon replacing one Mo with He.

Figure 3 displays the Friedel oscillations induced by one He atom in a (7×7×7) supercell Mo. The inset is a closeup of a narrower distance range. It has to be pointed out that by definition the Friedel oscillations are the relative electron density change in the metal with the introduction of an impurity. Thus, they are not meaningful right at the impurity atom, especially the core region. For this reason, we start the plot at $r$=0.2 $d_{nn}$, where $d_{nn}$ is the nearest-neighbor Mo-Mo distance. Clearly, oscillations in the <111> direction is the strongest, and that in the <110> direction which has the largest Mo-Mo distance is the weakest, indicating that chemical bonds have stronger response than interstitial itinerant electrons. Apparently, the first Mo atom along <111>, occupying the region $0.5 < r < 1.5$, loses electrons, the second Mo gains electrons, and the third also lose electrons, in an oscillation manner. The change in electron density beyond the third Mo is negligible. Along <100>, the first Mo atom, occupying the region $0.65 < r < 1.65$ lose electrons as in the <111> direction, the second Mo $1.8 < r < 2.8$ gain electrons yet with a much smaller magnitude, and the third Mo 2.95< $r$ < 3.95 undergoes a very slight electron lose. As for <110>, the first Mo $1.1 < r < 2.1$ gains electrons instead, in accordance with the fact that its distance to He is 1.6 $d_{nn}$, much larger than the nearest-neighbor. All these features explain extremely well the results of the calculated solution energies presented in Fig. 2 and provide a clear physical picture for the appearance of the high energy state around $r=2d_{nn}$, and low energy state around $r=3d_{nn}$ in the <111> direction. This implies that in the agglomeration process of implanted He atoms, there will be a remarkable potential barrier (0.7 eV) before the occurrence of He clustering, and there is a shallow potential well right before that barrier (0.1 eV).

**Fig. 4. Fitting of Friedel oscillations in <111> direction in Mo upon replacing one host atom with a He atom in a (7×7×7) Mo supercell. The insets are closeups for large *r*.**

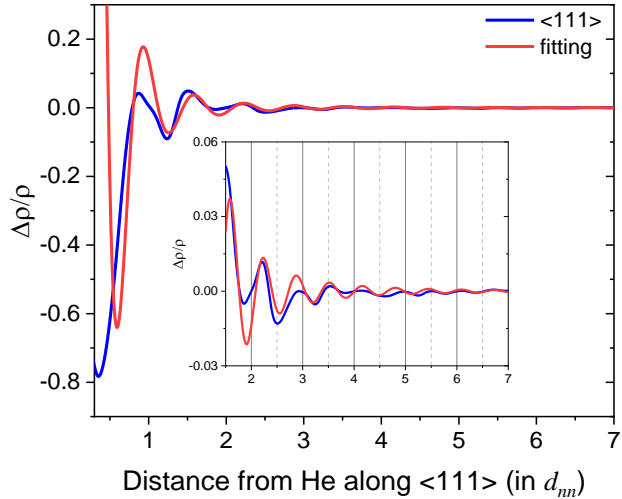

In Figure 4 we show the result of fitting for the Friedel oscillations in the <111> direction. Roughly, it has the following form

$$\frac{\Delta \rho}{\rho} = -0.15 \times \cos(9.8r) \times \frac{1}{r^3}. \qquad (3)$$

We note that Eq. (2) only holds for large $r$, so it is not unexpected that the He-induced Friedel oscillations in Mo do not follow closely the analytical form. The absence of a phase shift is mainly due to the chemical inertness of He, which prevents charge transfer between He and its surrounding atoms. The wavelength of oscillations is 0.64 $d_{nn}$, about 1.74Å, much smaller than that estimated from the Fermi velocity of Mo (0.92×10⁶ m/s), 7.9 Å. The reason for such a large discrepancy implies that when passing through atomic ions, the electron density variation in real metals cannot be describe properly by the jellium model.

Finally, we have also studied the He-He interaction at short range in the case of low concentration. We have varied the He-He distance in a (7×7×7) Mo supercell embedded with two substitutional He atoms sitting along the <111> direction. The calculated formation energy per He, as a function of He-He distance, is shown in Table 1. Very similar to the results displayed in Fig. 2 (solid and empty squares), we see clear oscillations of the He-He interaction at a fixed, low concentration 0.3 at%. The magnitude of oscillation is much smaller than in the high concentration case, as the number of nearest He neighbors is only two, while it is eight in the former case.

**Table 1. The formation energy, $\Delta H_s$ (eV/atom) of a He pair in <111> direction in a (7×7×7) Mo supercell at different Mo-Mo distance, $d$ (in scale of the nearest-neighbor distance in Mo, $d_{nn}$).**

| $d$ | 1 | 2 | 3 | 4 | 5 | 6 |
|---|---|---|---|---|---|---|
| $\Delta H_s$-fixed | 4.46 | 4.64 | 4.53 | 4.55 | 4.54 | 4.55 |
| $\Delta H_s$-optimized | 4.16 | 4.54 | 4.42 | 4.45 | 4.43 | 4.45 |

In summary, we have attempted to provide an understanding of the initial driving force for the formation of isomorphic He bubble superlattice in metals, taking Mo, in which this intriguing phenomenon was first observed, as an example. Realizing that the size mismatch between He and Mo is marginal and therefore the He-He elastic interaction should be very weak; we speculate that it is the He-induced Friedel oscillations of the electron density that produce the potential barriers to hinder the He clustering process. Our first-principles DFT calculations on the solution energy of ordered He atoms, which themselves form a superlattice in Mo, at different concentrations, demonstrate that at high concentration of 3.7 at% (=1/3$^3$), the isomorphic alignment of He (bcc) is energetically more favorable than fcc and simple cubic alignments. Moreover, an even higher concentration of 12.5 at% (=1/2$^3$) proves to be an obstacle for He to merge into clusters. Friedel oscillations are the macroscopic manifestation of the quantum nature of electrons. In this sense, the gas bubble ordering in metals serves as an excellent example of macroscopic quantum phenomenon.

Friedel had realized before the advent of electronic computers that charge density oscillations in metals induced by solute atoms could generate long-range interactions between them, but it is only with the help of high-performance supercomputers can we evaluate precisely such long-range effects. Our finding in this work proves unambiguously that Friedel oscillations can play important roles on mechanical properties in alloys, beside its well-known effect on electronic properties such as Knight shift in frequency of nuclear magnetic resonance (NMR) and Ruderman-Kittel-Kasuya-Yosida (RKKY) magnetic interactions in dilute alloys. We might expect the exploration of Friedel oscillations effect in alloys to deepen our understanding of solute-solute interactions which govern precipitation and decomposition processes, thereby promoting materials innovation.

**Acknowledgments**

This work was partially supported by the National Natural Science Foundation of China (No. 11775018).